\documentclass[useAMS,usenatbib]{mn2e}
\usepackage{graphicx}
\usepackage{txfonts} 

\newcommand{\aap}{A\&A}
\newcommand{\mnras}{MNRAS}
\newcommand{\apj}{ApJ}
\newcommand{\aj}{AJ}

\newcommand{\apjs}{ApJS}

\newcommand{\nat}{Nature}
\newcommand{\sci}{Science}

\catcode`\@=11
\def\gsim{\ifmmode{\mathrel{\mathpalette\@versim>}}
    \else{$\mathrel{\mathpalette\@versim>}$}\fi}
\def\lsim{\ifmmode{\mathrel{\mathpalette\@versim<}}
    \else{$\mathrel{\mathpalette\@versim<}$}\fi}
\def\@versim#1#2{\lower 2.9truept \vbox{\baselineskip 0pt \lineskip
    0.5truept \ialign{$\m@th#1\hfil##\hfil$\crcr#2\crcr\sim\crcr}}}
\catcode`\@=12
\def\msun{\hbox{$M_\odot$}}

\def\yr-1{\hbox{${\rm yr}^{-1}$}}

\def\t9{\hbox{$t_9$}}

\def\m*{\hbox{$M_{\rm stars}$}}

\def\ho{\hbox{$H_\circ$}}
\def\h50{\hbox{$\ho /50$}}

\begin{document}

\title{Disk Growth and Quenching} 
\author[Ying-jie Peng  and Alvio Renzini]{Ying-jie Peng$^{1,2}$\thanks{E-mail: yjpeng@pku.edu.cn}, and Alvio Renzini$^{3}$\thanks{E-mail: 
alvio.renzini@inaf.it}\\ 
$^{1}$Kavli Institute for Astronomy \& Astrophysics, Peking University, 5 Yiheyuan Road, Beijing 100871, China \\
$^{2}$Department of Astronomy, , School of Physics, Peking University, 5 Yiheyuan Road, Beijing 100871, China \\
 $^{3}$INAF - Osservatorio
Astronomico di Padova, Vicolo dell'Osservatorio 5, I-35122 Padova,
Italy}

\date{Accepted October 23, 2019; Received August 21, 2019, in original form}
 \pagerange{\pageref{firstpage}--\pageref{lastpage}} \pubyear{2002}

\maketitle
                                                            
\label{firstpage}

\begin{abstract}
Based on well established scaling relation for star forming galaxies as a function of redshift, we argue that the implied growth by a large factor of their angular momentum requires that the angular momentum of the inflowing gas fuelling star formation and disk growth must also secularly increase. We then propose that star formation in disks can cease (quench) once the accreted material (mainly atomic hydrogen) comes in with excessive angular momentum for sustaining an adequate radial flow of cold, molecular gas.
Existing observational evidence supporting this scenario is mentioned, together with some future observational studies that may validate (or invalidate) it.

\end{abstract}


\maketitle

\section{Introduction}
\label{intro}
By number, disk galaxies dominate as they do in terms of global star formation at least since redshift $\sim 2$. Hence, understanding their formation and evolution, especially in terms of stellar mass, structure and dynamics (including their angular momentum) has been a main theme in galaxy evolution studies starting with the seminal papers of \cite{peebles69}, \cite{doros70} and \cite{white84} that qualitatively predict a secular increase in the angular momentum of galaxies. Subsequent theoretical studies have been performed in the frame of the cold dark matter paradigm, with some success along with
some difficulties, arising from the complexity of the baryonic physics involved. Key contributions include  those of \cite{mo98}, \cite{ navarro00} and \cite{fall13} and more recently \cite{danovich15} and \cite{zoldan19}. 

Besides growing in mass, size and angular momentum, at a critical point in their evolution some disks cease to form stars, they quench, and continue evolving passively unless something happens that may re-ignite star formation. The physical origin of quenching remains matter of debate. In this paper we discuss in a semi-quantitative fashion disk growth and quenching basing uniquely on observational data, in particular  on galaxy scaling relations as a function of redshift as they have been established over the last decade or so, with the collective contribution of the broad astronomical community.

The paper is organised as follows. In Section 2 we very synthetically review the basic scaling relations for star forming galaxies. Section 3 deals with the implication
of the size and mass evolution of disks for the growth of their angular momentum and on the role of such growth for disk quenching. Indeed, we shall argue that the angular momentum of the gas inflow to galaxies can play a primary role not only on disk growth but also on their quenching. In Section 4 we mention some future observations that may help validating the proposed scenario and summarise its main aspects.

\section{The Galaxy Scaling Relations}

High redshift observations have revealed that by $z\simeq 2$ the majority ($\sim 70$ per cent) of star forming galaxies more massive than $\sim 10^{10}\,\msun$ have already settled into orderly rotating  disks, though with still high velocity dispersion compared to local disks
\citep{forster09,forster18,law09,glazebroock13,wisnioski15,simons17,ubler19}. This is also found in some recent hydrodynamical simulations (e.g., \citealt{pillepich19}). From such high redshift to the local Universe galactic disks evolve following --on average-- a set of empirically well established {\it scaling relations}, here briefly recalled. 

At fixed stellar mass, the half-light radius $R_{\rm h}$ of disks  scales as $\sim (1+z)^{-1}$ \citep{newman12,mosleh12,shibuya15,lilly16,mowla19}, so it increases by roughly a factor of $\sim 3$ over the last $\sim 10$ Gyr. This becomes a lower limit if one follows the evolution  of individual galaxies, as they growth also in mass and sizes increase with stellar mass, roughly as $R_{\rm h}\sim M_*^{0.2}$ \citep{vdw14,mowla19}. 
However, these figures refer to half-light radii, whereas half-mass radii appear to evolve slower, e.g., as $\sim (1+z)^{-0.5}$ \citep{mosleh17}, or even more so according to \cite{suess19}. Measured half-mass radii are still affected by large errors and different measuring procedures can result in very large differences for individual sources, though the distribution of the half-mass to half-light radii shows a sharp peak at 1, i.e., the two radii being equal  \citep{suess19}.

Disks at $z\sim 2$  have a much higher molecular gas fraction ($\sim 50$ per cent) than nearby disks \citep{tacconi10, daddi10,genzel15,scoville17}, which to first order scales as $\sim (1+z)^{2.6}$ \citep{tacconi18}.  Combining these two scaling relations we see that the surface gas density for fixed stellar mass scales as $\sim (1+z)^{4.6}$, which means that at $z\sim 2$  it is $\sim 150$ times (!) higher than in local disks of the same mass. 

Likely as a result of higher molecular gas content and gas density, the star formation rate (SFR) at fixed stellar mass increases as $\sim (1+z)^{2.8}$ (e.g., \citealt{speagle14,ilbert15}). Indeed, the close similarity of the two exponents of the molecular gas fraction and SFR scaling relations (that could actually be identical when allowing for the errors) imply that a Schmidt-like relation has been in place at least over the last 10 Gyr, and with a exponent close to unity (as in \citealt{bigiel08} for local galaxies and \citealt{tacconi18} at higher redshift). Of course, all of these scaling laws are affected by a sizeable dispersion, typically of the order of $\sim 0.2-0.4$ dex, implying that the evolutionary history of individual galaxies may differ substantially from the ideal case in which a galaxy strictly follows
all scaling relations.

Thus, for most of their lifetime star-forming galaxies appear to evolve following these scaling relations. In this framework, the ubiquity of galactic winds together with the need for continuous gas inflow to sustain star formation has led to the development simple toy models for the baryon cycle  through galaxies \citep{lilly13,peng14,dekel14a} in which near (yet not strict) equilibrium is maintained among the three rates of inflow, star formation and outflow. Moreover, a consistent  picture of the baryon cycle itself has become widely entertained, in which a bipolar outflow driven by supernova (and occasionally by AGN) feedback 
leaves the galaxy orthogonally from the disk, whereas inflow takes place predominantly in the equatorial plane, through the disk outer rim and corotating with the disk itself \citep{bouche13}. 

Direct evidence of this symmetry breaking has been recently demonstrated by \cite{crystal19a}  with a $z\sim 0.2$ sample of galaxy-quasar pairs in which the Mg II absorbing  circumgalactic medium (CGM) is found in corotation with the galaxy  in all 12 absorbers located within 46$^\circ$ of the major axis, whereas a nearly equal number of corotating and counterrotating absorbers is found at higher azimuthal angles, where outflows may dominate
(see also \citealt{ho19,crystal19b}).  Similarly, \cite{zabl19} find the Mg II absorbers being corotating in 7 out of 9 galaxies at $z\sim 1$. Thus, feeding galaxies by a mainly co-planar, corotating HI disk appears to be now a fairly well established paradigm, whereas 
extra-planar accretion was once favoured (see e.g., \citealt{sancisi08}, for a review). In summary, the disk-size scaling relation argument, hydrodynamical simulations  (e.g., \citealt{stewart11}) and direct observational evidence all concur in support of this scenario.

\section{The Growth and Quenching of Galactic Disks}
\subsection{Growth}
To a large extent, the necessity of this geometry (the equatorial accretion of mainly corotating material) follows simply from the $R_{\rm h} \sim (1+z)^{-1}$ scaling law: if disks grow in size, hence in angular momentum, then the new material must be added near the edge of the pre-existing disk and corotating with it, hence with higher and higher angular momentum as time goes by \citep{renzini18}.
The mere growth of disks requires that low-angular momentum gas is accreted first and high angular momentum gas later, with a long term (order of the Hubble time) coherence in remaining mainly planar and corotating. For, there would be little size and angular momentum growth if accretion were isotropic and with counterrotating and corotating gas flows being equally probable.

Also some hydrodynamical simulations hint to this scenario (e.g., \citealt{dekel09,stewart11,stewart17,danovich15,zoldan19}), but the emphasis here is on inferences we can draw directly from the observations. Thus, \cite{kassin12} find that from $z=1.2$ to $z=0$ the ratio of rotational velocity to gas velocity dispersion $V_{\rm rot}/\sigma_{\rm g}$ increases with time by the combined effect of an increasing $V_{\rm rot}$ and a declining $\sigma_{\rm g}$. Along the same track, \cite{simons17} find that at fixed stellar mass $V_{\rm rot}$ remains almost constant between $z\sim 2$ and the present for massive galaxies ($10<{\rm log}M_*/\msun<11$) but $V_{\rm rot}$ increases for individual galaxies as they grow in mass and conclude that galactic disks ``spin up as they assemble their mass". Again, this can only happen if the incoming material has higher angular momentum with respect to the already existing disk and symmetry between corotating and counterrotating inflows is broken in favour of the former ones.

On a more quantitative fashion, if at fixed stellar mass the radius evolves as $\sim (1+z)^{-1}$ and $V_{\rm rot}$  stays nearly constant, then also the stellar angular momentum evolves as $J_* \sim (1+z)^{-1}$, i.e., it must increase by a factor of $\sim 3$ from $z=2$ to the present. But in the same interval of cosmic time star forming galaxies increase in mass by a large factor that can be estimated by assuming that star-forming galaxies spend all their time on the Main Sequence (MS), hence the stellar mass growth comes from integrating the equation $dM_*/dt=(1-R)$SFR (e.g., \citealt{renzini09,peng10}), where $R\simeq 0.4$ is the mass return due to stellar mass loss. This mass increase is very sensitive to the adopted zero point and slope of the MS, such that between $z=2$ and $z=0$  it is a factor of $\sim 10$ for the MS adopted by \cite{speagle14} or a factor of $\sim 50$ for the steeper MS adopted by \cite{lilly13}, though in such case a galaxy may have quenched before growing so much. 

This mass increase implies  an additional factor $\sim 1.6$  increase in radius (the $R_{\rm h}\sim M_*^{0.2}$ scaling). Altogether,  by the combined effect of the mass and size increase, the angular momentum of individual galaxies would then increase by roughly a factor of $\sim 50$ between $z=2$ and the present (for the Speagle et al. MS), or evolving as $J_*\sim (1+z)^{-3.5}$, even neglecting the mentioned increase of $V_{\rm rot}$  experienced by  individual galaxies \citep{simons17}. This angular momentum increase is illustrated in Figure 1, which includes  the scaling $J_*\propto M_*^{5/3}(1+z)^{-1}$ proposed by \cite{swinbank17}.  This remarkable secular increase of the angular momentum of star forming disks is indeed a quite fundamental aspect of galaxy evolution. Note that a similar scaling law with redshift (hence cosmic time) needs to hold for the angular momentum of the inflowing material together with its long term corotation with the pre-existing disk (symmetry breaking).

This rate of angular momentum increase follows from the assumed $R_{\rm h}\sim (1+z)^{-1}$ scaling, but, as we have mentioned above,  the half-mass radius of star forming disks may evolve slower than this. For example, with $R_{\rm h}\sim (1+z)^{-0.5}$ (as e.g., in \citealt{mosleh17}), the total angular momentum of individual disks would increase by {\it only} a factor of $\sim 27$ from $z=2$ to the present. Thus, the precise scaling of the half-mass radii has an important effect on the inferred evolution of the angular momentum of disks, but the qualitative arguments developed here do not depend on the precise evolution of disk sizes, provided that a significant increase exists.

\begin{figure}
\vspace{-1.8 truecm}
\includegraphics[width=75mm,angle=-90 ]{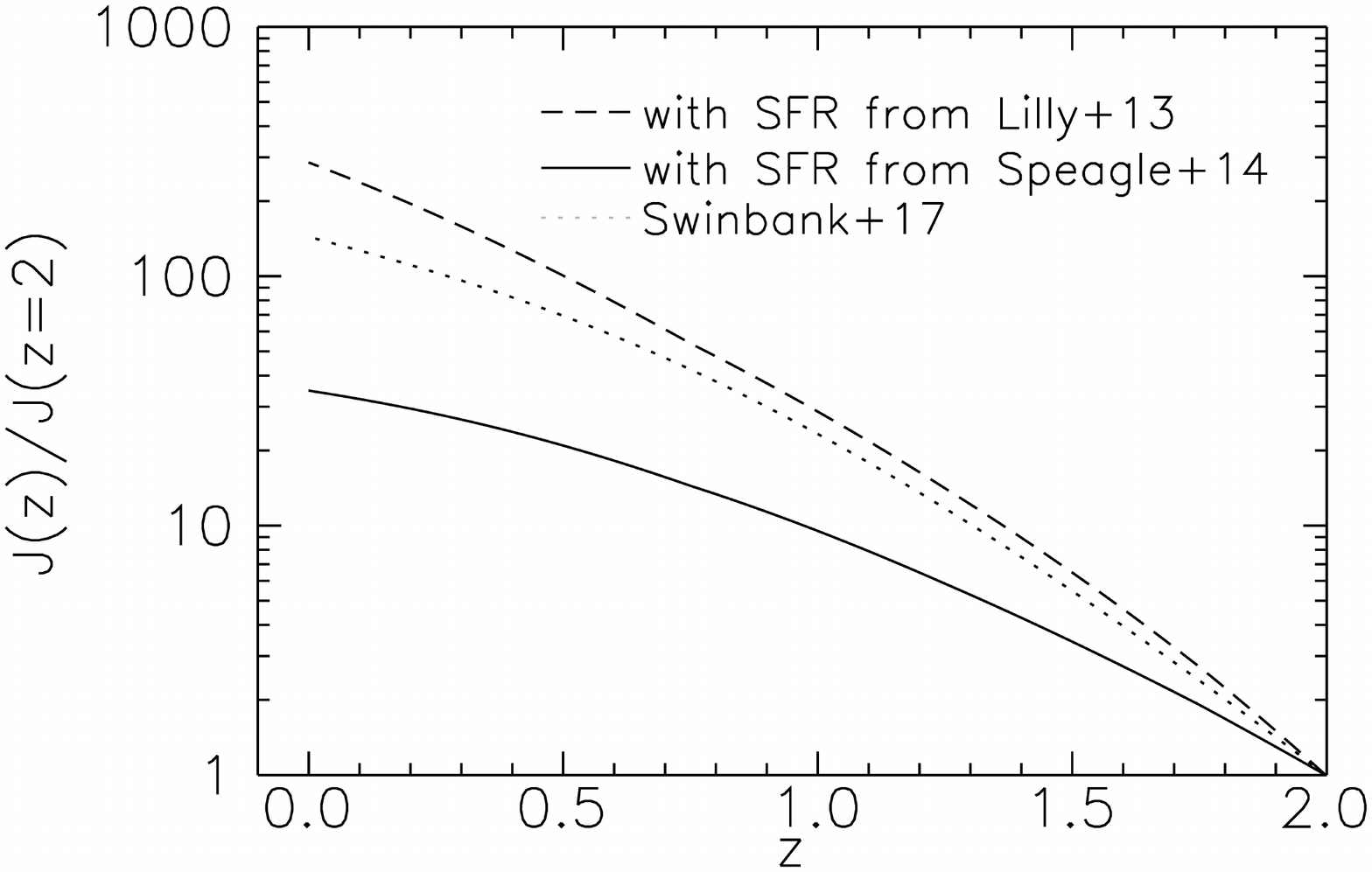}
\vspace{-0.8 truecm}
\caption{The average secular increase of the angular momentum of star-forming disks since $z=2$ for two choices of the SFR$(M_*,z)$ relation, from Lilly et al. (2013) and Speagle et al. (2014), as indicated. The angular momentum scaling from Swinbank et al. (2017) is also shown.}
\label{fig:mag_mass2}
\vspace{-0.8 truecm}
\end{figure}

\subsection{Quenching}

However, disks not only grow but at some point they can starve and eventually quench star formation altogether. Several quenching mechanisms have been proposed and the physics of quenching  remains a central issue in galaxy evolution studies. AGN feedback has been widely invoked, especially in semianalytic modelling, while coming in two possible versions: direct gas expulsion \citep{granato04}, then called {\it quasar-mode}, and heating of the CGM  by AGN jets \citep{croton06}, then called {\it radio-mode}. Alternative to AGN is the so-called {\it halo quenching} mechanism, in which CGM baryons get shock-heated to high temperatures when the host halo exceeds a threshold mass of $\sim 10^{12}\,\msun$ \citep{birnboim07}. Another option for quenching star formation in disks was  suggested by \cite{martig09}, called {\it morphological quenching}, in which the growth of a massive central bulge increases shear in the surrounding disk, thus preventing the formation of massive star-forming clumps. Moreover, ram pressure stripping can take place in galaxy groups and clusters \citep{gunn72}. Quenching may just occur as a result of rapid gas consumption by star formation and supernova driven polar outflows, such as in the hydrodynamical models of bulge formation following a violent disk instability, as in the hydrodynamical simulations of \cite{tacchella16}. Yet, the question remained how to maintain a bulge quenched while being still embedded in a star-forming disk. Based on the secular increase of the angular momentum of inflows, dictated by the scaling relations, \cite{renzini18} suggested  that a  bulge could remain quenched if the gas accreted via streams were to come in with too high angular momentum for reaching down to the bulge, which then would remain starved and almost completely passively evolving.

A recent finding about the HI content of central disk galaxies in the local Universe suggests that this quenching mechanism may not apply only to galactic bulges, but may even work for full galaxies, an ansatz we develop further here below. Indeed, using HI data, mostly  from the ALFALFA survey \citep{giovanelli05}, coupled to other databases, \cite{zhang19} were able to show that in the mass range $10.5<{\rm log}M_*/\msun<11$ the atomic hydrogen content of quenched disks is virtually identical to that of actively star-forming ones, whereas they lack of H$_2$ molecular gas. Moreover, the quenched central disks exhibit the characteristic double-horn HI profiles as 
star-forming disks, suggesting that both galaxy types have regularly rotating HI disks. The almost constant HI gas mass of $10^{10}\,\msun$ from star-forming to quenched disks corresponds to HI disks of radius about 30 kpc according to the HI size-mass relation \citep{broeils97,wang16}.Though the radio data do not have enough  spatial resolution, it is most likely that the HI gas in quenched disks lies in a ring external to the optical disk, orbiting with too high angular momentum to be able to spiral in, compress, undergo a  transition to the H$_2$ molecular phase and eventually turn into stars.  HI ring-like structures with central HI hole are indeed found in several galaxies and in particular in quenched disks \citep[e.g.,][]{murugeshan19}.

What would then distinguish star-forming disks from quenched ones, if both have the same HI content? One possibility is that in quenched disks the outer HI ring  is on stable orbits and relatively unperturbed whereas in star-forming disks the HI ring may be occasionally, or even continuously,  perturbed by encounters with other galaxies, or by galactic satellites, such as in the case of the Magellanic Clouds for the Milky Way or M32 and NGC205 for Andromeda. Such disk perturbations could then induce deviations from circular orbits, hence radial flows that can promote and sustain star formation. Temporary reactivation of radial inflow and star formation could also result from an occasional  counterrotating stream or a minor merger. In any event, radial transport of gas is considered a requisite to sustain star formation, that is, to avoid quenching \citep{krumholz18}.

Hence, we here propose angular momentum inhibition of radial flows as an additional quenching mechanism, specific for disk galaxies. In practice,  disk growth and its quenching are seen as being the result of the same process, namely accretion by  planar, corotating inflows with secularly increasing angular momentum that ensure disk growth until the cold, atomic gas eventually settles in too high angular momentum orbits for being able to further penetrate inside and promote star formation.
Indeed, the direct evidence from HI-rich quenched disks is that these galaxies are indeed fed (by HI) but they don't {\it eat}, i.e., they are unable to process atomic gas to molecular clouds and stars. So they starve and quench.
Yet, this kind of quenching may not be definitive, as the arrival of an external perturbation, be it a minor merger or a counterrotating stream, may shake the HI reservoir, inducing radial inflows and revitalising star formation, thus {\it rejuvenating} the once quenched galaxy (e.g., \citealt{saintonge11,mancini19}).

One question that needs to be answered is why the HI ring does not spontaneously form molecular clouds and stars. The answer to this question comes from a recent study by \cite{obres16} showing that disks quite naturally develop a positive radial gradient in Toomre's stability parameter $Q$ and in the local Universe the gas in the outer regions indeed have a large $Q>1$, hence should be stable against fragmentation and remain atomic. Alternatively, but this may be part of the just mentioned effect, the HI in the outskirts of (quenched) disks may be at densities below the empirical threshold above which HI is converted to H$_2$ ($\sim 10\;\msun\, {\rm pc}^{-2}$, \citealt{bigiel08}). Very low metallicity could also inhibit grain, hence H$_2$ formation, yet, the corotating gas is revealed by MgII absorption, hence it contains metals.

Though in our imagination the prototypical quenched galaxy is a giant elliptical, the bulk of first quenched galaxies are actually disks \citep{bundy10,vdw11,toft17,bezanson18,newman18} and so are the bulk of quenched, early-type galaxies in the local Universe that are still fast rotators \citep{emsellem11}. So, it is the quenching of disks that dominates quenching in general, and this proposed mechanism to starve disks by feeding them gas they cannot reach has the attractive aspect of being germane to disks and their growth, as it is the growth itself that prepares the conditions for the subsequent quenching. In a sense, disks quench when they have consumed all the relatively low angular momentum gas.We further venture to speculate that this kind of quenching  would very naturally affect first the most grown galaxies, reminiscent of the \cite{peng10} {\it mass quenching} paradigm. Yet, it remains to demonstrate that this starvation by excessive angular momentum  starts working just once the stellar mass has reached close to the {\it Schechter mass}, as in the mass quenching mode of \cite{peng10}\footnote[4]{Note, however, that these aspects, growth preparing for quenching and why the Schechter mass has the  observed value,  are common to other quenching mechanisms as well.}. This kind of quenching may also be relatively slow, with a great deal of variance from one galaxy to another, and being basically a quenching by {\it strangulation} as in  the slow-quenching mode proposed by \cite{peng15}. Moreover, in this scenario quenching is not accompanied by a morphological transformation, as indeed suggested by freshly quenched galaxies at high redshift being still fast rotating disks \citep{toft17,newman18}.

Another attractive aspect of quenching by excessive angular momentum is that it quite naturally leads to inside-out quenching for which evidence exists both at high redshifts (e.g., \citealt{tacchella15,vkk15,spilker19}) and in the local Universe \citep{guo19,morselli18}. This is particularly evident in the stacks of mass and SFR surface density performed in these last two studies, in which the central (bulge) regions of galaxies are either fully quenched or have depressed specific SFR, whereas star formation persists in the outer regions, though with very low rates in quenched galaxies. Exception to this trend are starburst galaxies, in which the SFR surface density actually peaks at the centre \citep{guo19,morselli18}. 

Of course, this  mode does not exclude that other quenching mechanisms may also be at work. However, it seems to us that some of the traditional quenching mechanisms cannot be invoked to account for disks that are quenched in spite of there being still plenty of atomic hydrogen in their reservoir. 
We finally emphasise, once more, that  all this ultimately follows from just the $R_{\rm h} \sim (1+z)^{-1}$ scaling relation.

\section{Possible Observational (In)Validations}
As mentioned earlier, the ALFALFA 21 cm data do not have sufficient resolution for determining the structure of the HI emitting gas. It would then be important to verify whether in quenched disks the HI material is indeed in a ring located outside the optical radius, as we have assumed above.

Another question concerns what makes the difference between star-forming and quenched disks, in spite of them having comparable HI content. This may have to do with the geometry of the HI distribution and/or the presence/absence of other objects (companions, satellites, etc.) capable of destabilising the HI ring thus inducing  radial inflow. For example, do massive, quenched disks inhabit preferentially low density environments and/or are they satellite-less centrals?  Or, compared to star-forming central disks, do they have different structure in terms of disk size, concentration, stellar mass
surface density, bar frequency or AGN activity? We will explore all these issues in a future paper.

Current information on HI gas in galaxies is basically limited to the local Universe. However, with the advent of SKA the situation will change dramatically. For example, the Ultra-deep SKA1 surveys will probe massive galaxies (with $M_{\rm HI}\gsim  10^{10}\,\msun$)  all the way to  $z\lsim 1.7$ \citep{blyth15}, hence mapping the evolution of the HI content of galaxies (both quenched and star-forming) over a major fraction of cosmic time, thus complementing the molecular gas measurements as from CO, C$^+$ and dust continuum. 

In this respect, it would be interesting to extend to quenched disks at high redshift the traditional absorption studies in the spectra of nearby quasars, a combination that may have been neglected compared to star-forming disks.

Determining the redshift evolution of the half-mass radii of galaxies more precisely than currently available would greatly help to put this scenario on firmer quantitative grounds. If there were no or very little evolution, the proposed quenching by secular angular momentum increase could be invalidated. Moreover, of critical importance is to most accurately establish the  scaling relation for the stellar angular momentum of galaxies.

Direct observational evidence of corotating gas inflows and and fast-rotating quenched disks is currently limited to few objects whereas both should dominate for the proposed quenching mechanism to hold. Therefore, a crucial test would come from a substantially increased statistics.

In this paper we have argued that the observed secular increase in the size, stellar mass and rotational velocity of galactic disks implies a remarkable secular increase in their angular momentum that for individual galaxies can exceed a factor $\sim 30$ since redshift $\sim 2$. We then propose a new  mechanism for the quenching of star formation in galactic disks, that would arise from the gas inflow into disks eventually coming in with too high angular momentum for continuing to supply molecular gas to the inner regions of galaxies. The incoming gas would instead settle on an outer ring of neutral hydrogen, stable against fragmentation and radial migration, as indeed observed in quenched galactic disks in the local Universe.

One interesting aspect worth further study is why the characteristic mass above which the quenching probability grows exponentially, equal to the Schechter mass of star forming galaxies \citep{peng10}, is nearly constant with redshift up to $z\sim 2$, whereas the size of quenched galaxies strongly decreases with redshift. This is an issue that hydrodynamical simulations may help to understand.

\section*{Acknowledgments}

We would like to thank Natascha F\"oster Schreiber, Mauro Giavalisco and Giulia Rodighiero for stimulating discussions on these subjects.
AR acknowledge support from an INAF/PRIN-SKA 2017 (ESKAPE-HI) grant.  YP acknowledges the National Key R\&D Program of China, Grant 2016YFA0400702 and NSFC Grant No. 11773001.
% MITIC ??

\label{lastpage}

\end{document}